\documentclass[aps,prl,twocolumn,floatfix,showpacs]{revtex4}
\usepackage{graphicx} 
\usepackage{floatrow}
\usepackage{ulem}
\usepackage{color}
\usepackage{enumerate}
\usepackage{enumitem}
\usepackage[colorlinks=true,linkcolor=blue,citecolor=blue,urlcolor=blue]{hyperref} 
\begin{document}
\title{Origin of second harmonic generation in non-centrosymmetric crystal structures containing lone-pairs electrons}
\author{Fuming Li$^{1,2}$}

\affiliation{$^{1}$Research Center for Crystal Materials; CAS Key Laboratory of Functional Materials and Devices for Special Environmental Conditions; Xinjiang Key Laboratory of Functional Crystal Materials; Xinjiang Technical Institute of Physics and Chemistry, Chinese Academy of Sciences, Urumqi 830011, China}
\affiliation{$^{2}$Center of Materials Science and Optoelectronics Engineering, University of Chinese Academy of Sciences, Beijing 100049, China}

\author{Shilie Pan$^{1,2}$}
\email{slpan@ms.xjb.ac.cn}
\affiliation{$^{1}$Research Center for Crystal Materials; CAS Key Laboratory of Functional Materials and Devices for Special Environmental Conditions; Xinjiang Key Laboratory of Functional Crystal Materials; Xinjiang Technical Institute of Physics and Chemistry, Chinese Academy of Sciences, Urumqi 830011, China}
\affiliation{$^{2}$Center of Materials Science and Optoelectronics Engineering, University of Chinese Academy of Sciences, Beijing 100049, China}

\author{Zhihua Yang$^{1,2}$}
\email{zhyang@ms.xjb.ac.cn}
\affiliation{$^{1}$Research Center for Crystal Materials; CAS Key Laboratory of Functional Materials and Devices for Special Environmental Conditions; Xinjiang Key Laboratory of Functional Crystal Materials; Xinjiang Technical Institute of Physics and Chemistry, Chinese Academy of Sciences, Urumqi 830011, China}
\affiliation{$^{2}$Center of Materials Science and Optoelectronics Engineering, University of Chinese Academy of Sciences, Beijing 100049, China}

\begin{abstract}
Material systems with lone-pair electrons have long been a treasure trove in the search for large second harmonic generation effects. Revealing the origin of second harmonic generation in nonlinear optical materials can provide theoretical guidance for the design of new materials. In this work, the origin of second harmonic generation in non-centrosymmetric materials containing lone pair electrons is revealed by analyzing the orbital interactions on the sublattice. Stereochemically inactive Pb 6\textit{s} orbitals with high symmetry in CsPbCO$_3$F contribute less to the second harmonic generation. In contrast, the contribution of stereochemically active Pb 6\textit{s} orbital in PbB$_5$O$_7$F$_3$ and PbB$_2$O$_3$F$_2$ is more obvious. Significantly, the orbitals of the interaction between lead and oxygen make a very significant contribution because these orbitals are located at the band edge and in non-centrosymmetric sublattices.
\end{abstract}


\maketitle

\section{INTRODUCTION}

Nonlinear optical (NLO) effects play an important role in fundamental scientific research and the practical application of high-performance optoelectronic materials \cite{R01,R02,R03,R04,R05,R06}. NLO materials exhibit an extraordinary ability to convert laser frequencies \textit{via} second harmonic generation (SHG). The polarization dependence of SHG can provide rich information about the structural symmetry of the material and the microscopic arrangement of chemical bonds \cite{R07,R08,R09,R10,R11}. The atomic orbitals in materials form hybrid orbitals with diverse shapes and energies, which reflects the characteristics of chemical bonds and molecular geometry \cite{R12}. Generally, the chemical-bond-associated microstructures determine the physico-chemical properties of materials, which are inseparable from the covalent interaction (\textit{e.g.}, B/P-O bonds) of microscopic groups or driven of electrostatic ionic systems (\textit{e.g.}, M$_x$O$_y$ polyhedron, M is for metal elements) \cite{R13,R14}. For example, depending on the asymmetric charge distribution of $\pi$ orbital itself, the material may have a large SHG effect \cite{R15}. For materials with $\textit{ns}^2$ lone-pair electron cations such as Pb$^{2+}$, Bi$^{3+}$ and I$^{5+}$, the interaction of cation \textit{s}-state and anion \textit{p}-state orbitals can affect the strength of chemical bonds and deform the distortion of coordination geometry \cite{R16}, which gives rise to some interesting and fascinating physical properties \cite{R17,R18}. For example, it plays an important role in both the exciton relaxation and dissociation of low-dimensional halide perovskites \cite{R19}. Lone pair also can increase phonon dissonance, resulting in low lattice thermal conductivity \cite{R20}. In NLO materials, lone-pair electron cations may improve the microscopic second-order polarizability of ionic groups, thereby inducing greater NLO properties \cite{R21,R22,R23}. Reasonably assembling polar units can be used to effectively design non-centrosymmetric materials with excellent SHG performance \cite{R24}. For example, the IO$_3$ group belongs to the lower symmetry $C_{3v}$ group compared with BO$_3$ ($D_{3h}$) own to the lone-pair electrons on the iodine cation and has a large microscopic second-order polarizability \cite{R25}.  The SHG effect can be enhanced by condensing IO$_3$ in a parallel arrangement in each unit cell, such as \textit{\text{$\alpha$}}-LiIO$_3$ crystal \cite{R26}. Therefore, material systems with lone-pair electrons have long been a treasure trove in the search for large SHG effects.

Lone-pair electrons can exhibit varying degrees of stereochemical activity or even inactivity, which is governed by crystal symmetry and orbital energy differences \cite{R16}. The asymmetric distribution of electron clouds at the band edge indictates that the stereochemically active lone-pair electrons can induce excellent optical properties of materials. For example, SbB$_3$O$_6$ contains the stereochemically active lone-pair Sb$^{3+}$ cation, which gives it the highest birefringence (0.290 at 546 nm) among borate materials, and also exhibits a strong SHG effects of 3.5 $\times$ KH$_2$PO$_4$ (KDP) \cite{R27}. On the other hand, significant SHG effects can also be enhanced by \textit{p}-$\pi$ interactions of cations and $\pi$-conjugated groups in materials containing stereochemical inactivity, \textit{e.g.}, CsPbCO$_3$F \cite{R28}. Understanding the structure-property relationships of these materials is crucial for elucidating the effect of lone-pair electrons on SHG. The wave function analysis under local representation in periodic lattice potential field provides an effective method to study the factors affecting the optical properties of materials \cite{R29}. Currently, the influence of lone-pair electrons with varying degrees of stereochemical activity on the SHG effects has not been directly investigated at the orbital level. Specifically, quantitative studies are still needed regarding the direct contribution of cations with $\textit{ns}^2$ lone-pair electrons to the SHG coefficient and the influence of the orbitals through which these cations interact with surrounding anions. In this work, we investigated the origin of SHG in materials containing different stereochemically active lone-pair electrons.

{\section{COMPUTATIONAL ANALYSIS METHOD}}
All first-principles simulations were performed using Vienna \textit{ab} initio simulation package (VASP) \cite{R30}. The exchange-correlation potential and ion-electron interaction were described by the Perdew-Burke-Ernzerhof (PBE) functional and projector-augmented wave (PAW) methods, respectively. The valence electron configurations were treated as follows: Cs-\textit{s},\textit{p},\textit{d},  Pb-\textit{s},\textit{p},\textit{d}, Sn-\textit{s},\textit{p},\textit{d}, B-\textit{s},\textit{p}, C-\textit{s},\textit{p}, O-\textit{s},\textit{p} and F-\textit{s},\textit{p}. First, the Broyden-Fletcher-Goldfarb-Shannon (BFGS) algorithm is used to optimize the atomic position and lattice parameters \cite{R31,R32}, and the atoms are allowed to relax until the force applied to the atom is less than 0.02 eV/\AA. An energy cutoff of 400 eV and specific Monkhorst-Pack \textit{k}-point meshes of \(7 \times 7 \times 7\) for CsPbCO\(_3\)F, \(3 \times 4 \times 4\) for PbB\(_5\)O\(_7\)F\(_3\), and \(9 \times 9 \times 7\) for PbB\(_2\)O\(_3\)F\(_2\) were used. Second, during the static self-consistent-field calculation, the plane-wave cutoff energy of 600 eV and the threshold of 10$^{-7}$ eV, where the \textit{k}-point meshes is consistent with the above. Third, the Wannier orbitals are constructed through a post processing procedure using the output of VASP calculation, and the corresponding orbitals type by the projection of all valence states in the unit cell were generated using Wannier90 \cite{R33,R34,R35}. And the plane-wave cutoff energy of 800 eV, threshold of 10$^{-10}$ eV and dense Monkhorst Pack k-point mesh spanning remains the same. Fourth, the optical properties of all compounds and the SHG contribution of each Wannier orbital were calculated. For the optical property calculations, the plane-wave cutoff energy of 800 eV, energy threshold of 10$^{-10}$ eV, and a Monkhorst-Pack k-point mesh with a density twice that used in the self-consistent-field calculation were applied to ensure higher precision.

\begin{figure}[H]
	\centering
	\includegraphics[width=1\textwidth]{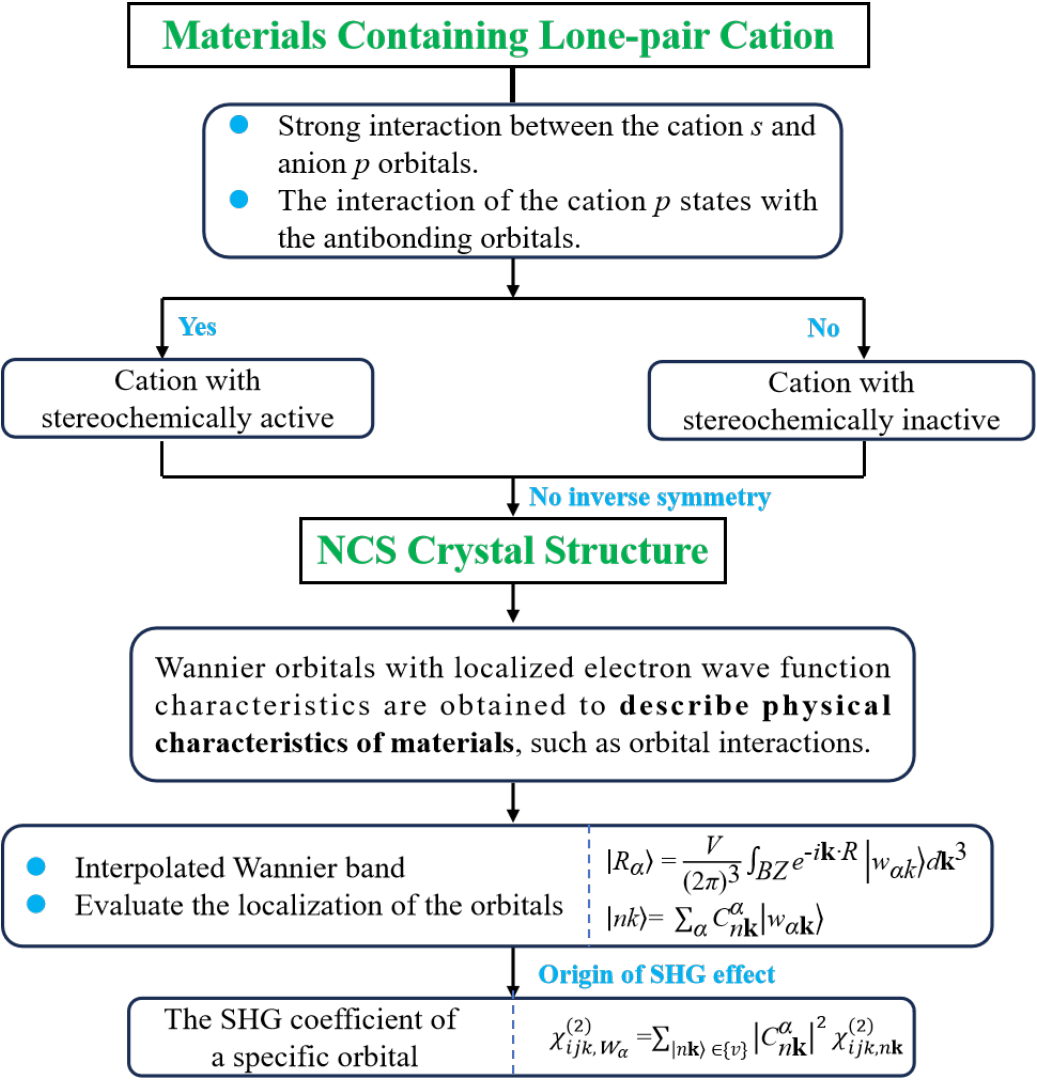} 
	\caption{(Color online) Analysis of the decision tree for the origin of SHG coefficients of materials containing lone-pair cation.}
	\label{fig:decision_tree}
\end{figure}

The length gauge formalism method, originally developed by Aversa and Sipe \cite{R36}, has proven effective in avoiding unphysical divergences that may arise in NLO calculations. Rashkeev et al.\cite{R37} rearrange their formula, and its static frequency doubling coefficient calculation formula can be further simplified, the frequency dependent form derived from the length gauge method can be simplified to the form derived from the velocity gauge method \cite{R38}. At zero frequency, the SHG coefficients $\textit{d}_{im}$ can be derived by transforming from the second-order susceptibility tensor $\chi_{ijk}^{(2)}$, where the relationship is given as $\textit{d}_{im} = \frac{1}{2}\chi_{ijk}^{(2)}$. Specifically, they are calculated using the following expression:

\begin{equation}
	\label{eq:NN_def}
		\chi_{ijk}^{(2)}=\chi_{ijk}^{(2)}(VE)+\chi_{ijk}^{(2)}(VH),
\end{equation}
\onecolumngrid
	\begin{equation}
		\label{eq:NN_def}
			\chi_{ijk}^{(2)}(\mathrm{VE})= \frac{e^3}{2\hbar^2m^3}\sum_{vcc^{\prime}}\int\frac{d\mathbf{k}^3}{4\pi^3}P(ijk)\mathrm{Im}[p_{vc}^ip_{cc^{\prime}}^jp_{c^{\prime}v}^k] \\
			\left(\frac1{\omega_{cv}^3\omega_{vc^{\prime}}^2}+\frac2{\omega_{vc}^4\omega_{c^{\prime}v}}\right),
	\end{equation}

    \begin{equation}
	\label{eq:NN_def}
		\chi_{ijk}^{(2)}(\mathrm{VH})= \frac{e^3}{2\hbar^2m^3}\sum_{vv^{\prime}c}\int\frac{d\mathbf{k}^3}{4\pi^3}P(ijk)\mathrm{Im}[p_{vv^{\prime}}^ip_{v^{\prime}c}^jp_{cv}^k] \\
	 	\left(\frac1{\omega_{cv}^3\omega_{v^{\prime}c}^2}+\frac2{\omega_{vc}^4\omega_{cv^{\prime}}}\right),
    \end{equation}
\twocolumngrid

The \textit{i}, \textit{j}, \textit{k} are Cartesian components, \textit{v} and \textit{v}${'}$ denote valence bands, \textit{c} and \textit{c}${'}$ denote conduction bands, and \textit{P}(\textit{ijk}) denotes full permutation and explicitly shows the Kleinman symmetry, which ensures the frequency-independent nature of the nonlinear susceptibility tensor. In addition, the contribution of the two-band transition process has been strictly proved to be zero through rigorous theoretical analysis \cite{R39}. It is worth noting that Wannier90 has integrated this functionality using the length gauge formalism, and the calculation results of the SHG coefficient are fits well with our current method \cite{R40}.

Figure \ref{fig:decision_tree} presents a decision tree framework for analyzing the origin of SHG coefficients in crystalline materials. For specific materials, the analysis first evaluates the stereochemical activity of lone-pair electrons by examining two key factors: the local symmetry around the cation and the orbital energy alignment between cation and anion orbitals. Then, by combining information about the crystal symmetry (centrosymmetric or non-centrosymmetric) with the degree of stereochemical activity of lone pairs, their influence on the material microstructure and SHG response can be further analyzed.

Using a unitary transformation, a set of Wannier functions (WFs) \(\textbf{\textit{R}}(\textit{r})\), labeled by the Bravais lattice vector \(\textbf{\textit{R}}\), can be constructed from the Bloch eigenstates \(|n\textbf{k}\rangle\) of band \(n\). This transformation preserves all the physical information while providing a more intuitive real-space picture of the electronic structure. To quantitatively evaluate the contribution of Wannier orbitals to the SHG effects, a projection procedure is implemented. All valence states \( |n\mathbf{k}\rangle \in \{v\} \) (where \(\{v\}\) represents the set of all valence bands) are projected onto atomiclike Wannier orbitals \( |w_\alpha\rangle \). Through this projection, we obtain the coefficients \( C_{nk}^\alpha \), which quantify how much each atomiclike Wannier orbital contributes to each valence state. The square of the magnitude of these projection coefficients, \( |C_{nk}^\alpha|^2 \), represents the weight or contribution of the Wannier orbital \( w_\alpha \) to the \(n\)-th valence band-decompositon \cite{R41}. This decomposition allows us to directly evaluate the orbital-resolved contributions to the SHG effects, providing detailed insights into the microscopic origin of NLO properties \cite{R42}.
\vspace{1\baselineskip}
{\section{RESULTS AND DISCUSSION}}

\begin{figure}[!b]
	\centering
	\includegraphics[width=1\textwidth]{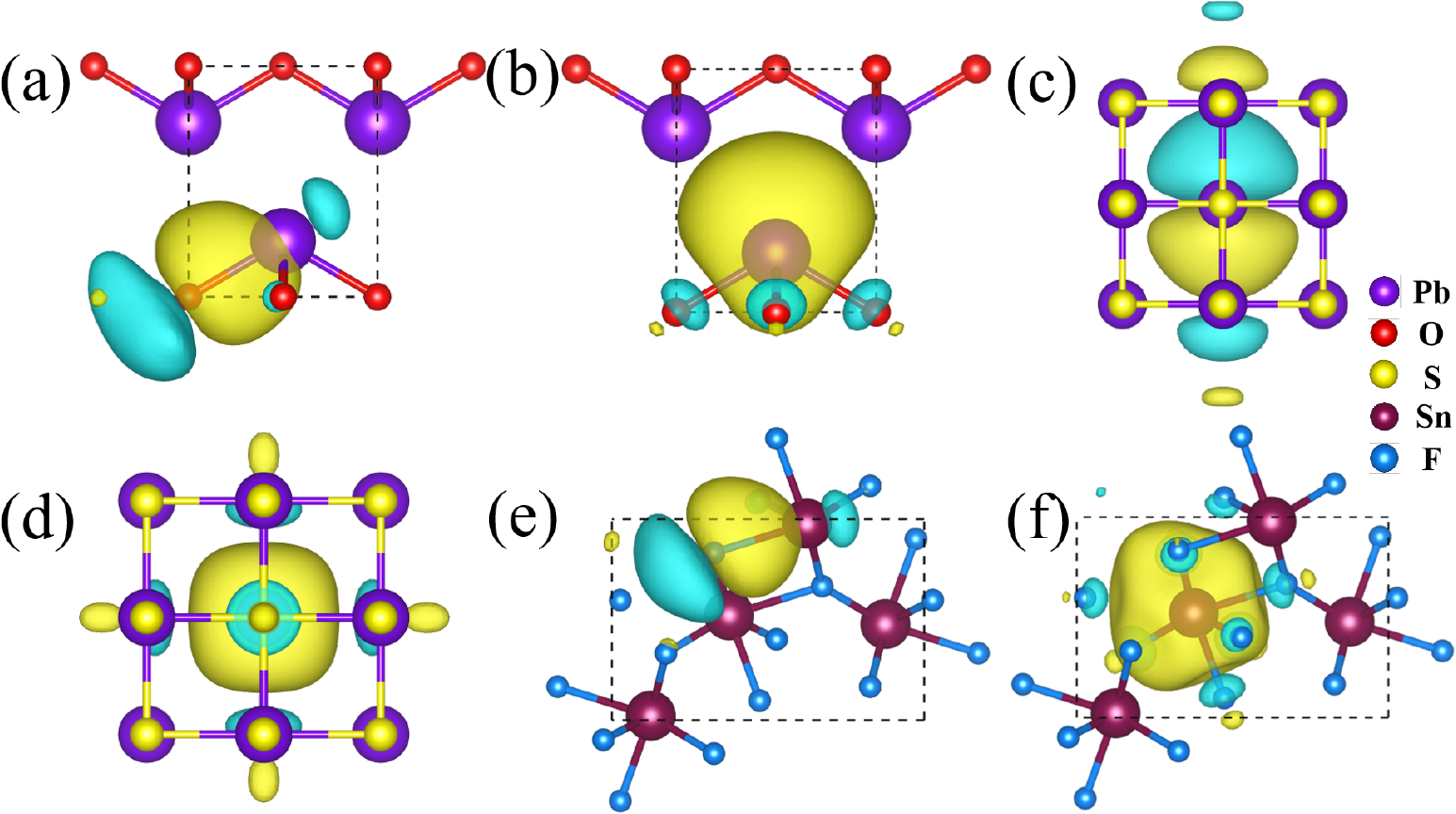} 

	\caption{(Color online) The Pb-O ionic orbital (a) and Pb 6\textit{s} orbital (activity) (b) in litharge PbO. The Pb-S ionic orbital (c) and Pb 6\textit{s} orbital (inactivity) (d) in rocksalt PbS. The Sn-F ionic orbital (e) and Sn 5\textit{s} orbital (f) in \textit{\text{$\beta$}}-SnF$_2$.}
	\label{fig:wannier_orbitals}
\end{figure}
To illustrate the types of stereochemical activity and inactivity lone-pair electron at different sublattice and their effects on electronic structural distortion, we first describe the orbital interactions in the valence bands of litharge PbO and rocksalt PbS \cite{R43,R44,R45}. The interpolated bands of maximally projected WFs are shown in dotted lines in Figs. S1a-S1b, which matches the  original band structure (solid line), indicating that well localized of the Wannier orbital. Here, the Wannier orbital is a linear combination of all relevant atomic orbitals in energy space. Interactions between these atomic orbitals cause the shape of WFs to deviate from the standard atomic orbitals. Figures \hyperref[fig:wannier_orbitals]{\ref{fig:wannier_orbitals}(a)--\ref{fig:wannier_orbitals}(d)} show the Wannier orbital of lead and the oxygen/sulfur-centered for PbO and PbS. The features of oxygen orbitals expand in the direction of lead and have covalent characteristics with lead consist of Pb 6\textit{s}/6\textit{p} and O 2\textit{p} [see Fig. \hyperref[fig:wannier_orbitals]{\ref{fig:wannier_orbitals}(a)}]. The differences of Pb 6\textit{s} orbitals can be observed, which shows the activity and inactivity of lone-pair electrons in PbO and PbS, respectively. In PbO, a highly asymmetric lobe is away from lead, which is determined by the anti-bonding (Pb 6\textit{s} + O 2\textit{p})* and Pb 6\textit{p}. In contrast, lead in PbS is constrained by the symmetry of surrounding sulfur atoms, the Pb 6\textit{s} orbitals exhibit a completely symmetrical shape and will not exhibit stereochemical activity. The litharge PbO and rocksalt PbS structure locate in a centrosymmetric space group, so all orbitals forming centrosymmetric sublattices and do not contribute to SHG. In contrast, \textit{\text{$\beta$}}-SnF$_2$ is located in a chiral \textit{P}4$_1$2$_1$2$_1$ space group, and similar methods were used to analyze orbital interactions (Fig. \hyperref[fig:wannier_orbitals]{\ref{fig:wannier_orbitals}(e)--\ref{fig:wannier_orbitals}(f)}). The Sn 5\textit{s} orbital of \textit{\text{$\beta$}}-SnF$_2$ also exhibits an asymmetric electronic distribution, and the symmetry analysis shows that the Sn 5\textit{s} orbital is located in a non-centrosymmetric sublattice, which dominates the contribution to the SHG.

\begin{figure}[!b]
	\centering
	\includegraphics[width=1\textwidth]{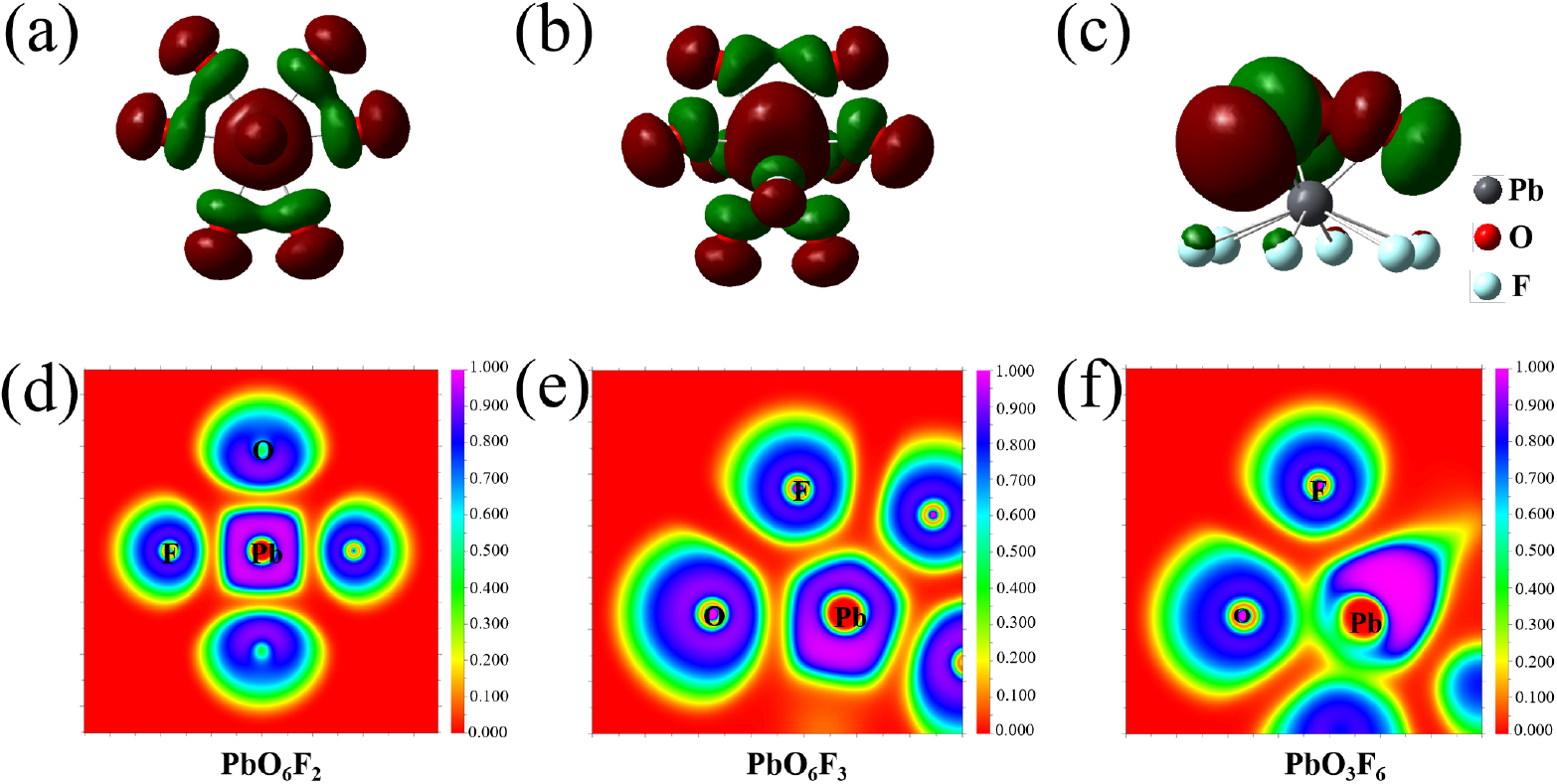} 
	\caption{(Color online) The highest occupied molecular orbital (a-c) and ELF diagram (d-f) of PbO$_6$F$_2$, PbO$_6$F$_3$ and PbO$_3$F$_6$ in CsPbCO$_3$F, PbB$_5$O$_7$F$_3$ and PbB$_2$O$_3$F$_2$.}
	\label{fig:homo}
\end{figure}

\begin{figure*}[!t]
	\centering
	\includegraphics[width=0.93\textwidth]{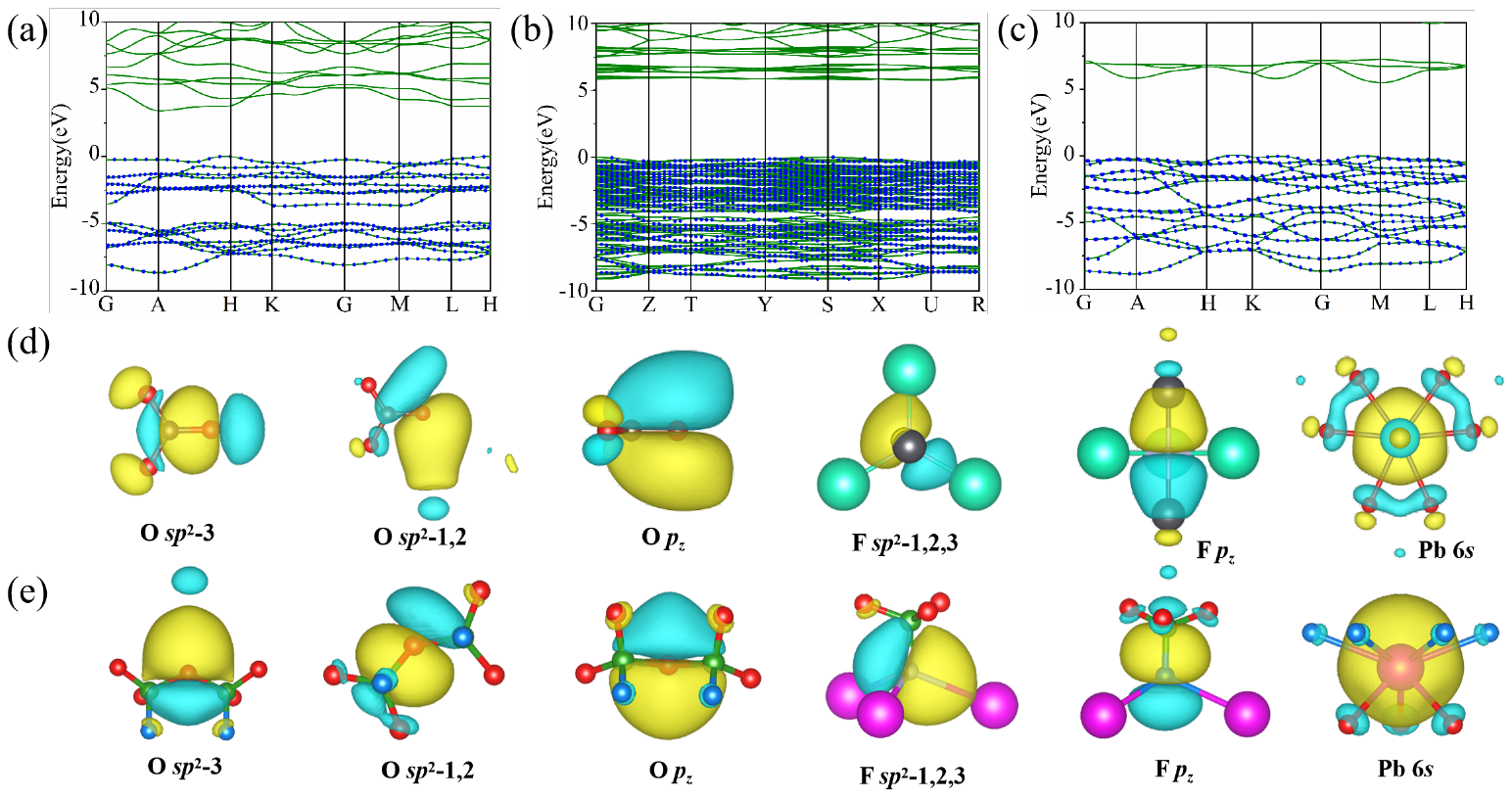} 
	\caption{(Color online) Band structure (solid lines: DFT bands; dots: Wannier-interpolated bands) of CsPbCO$_3$F (a), PbB$_5$O$_7$F$_3$ (b) and PbB$_2$O$_3$F$_2$ (c). Wannier orbitals of oxygen, fluorine and lead in CsPbCO$_3$F (d): from left to right are: the orbital of \textit{$\sigma$} bonds between carbon and oxygen (O \textit{sp}$^2$-3); the orbital of ionic interaction between oxygen and lead (O \textit{sp}$^2$-1,2); nonbonding orbital of oxygen (O \textit{p}$_z$); the orbital of ionic interaction between fluorine and cesium (F \textit{sp}$^2$-1,2,3); the orbital of ionic interaction between fluorine and lead (F \textit{p}$_z$); and Pb 6s orbital. Wannier orbitals of oxygen, fluorine and lead in PbB$_2$O$_3$F$_2$ (e): the orbital of ionic interaction between oxygen and lead (O \textit{sp}$^2$-3); the orbital of \textit{$\sigma$} bonds between boron and oxygen (O \textit{sp}$^2$-1,2); nonbonding orbital of oxygen (O \textit{p}$_z$); the orbital of ionic interaction between fluorine and lead (F \textit{sp}$^2$-1,2,3); the orbital of \textit{$\sigma$} bonds between fluorine and boron (\textit{p}$_z$); and Pb 6\textit{s} orbital.
	}
	\label{fig:band_structure}
\end{figure*}

To identify the origin of SHG in materials with different stereochemical activity materials at the orbital level, we analyzed the SHG properties of CsPbCO$_3$F, PbB$_5$O$_7$F$_3$ and PbB$_2$O$_3$F$_2$ by examining their lattice symmetries and electronic structures. Here, we first divided the sublattices of the crystal structure of these materials, which can be decomposed into different chemical environment sublattices containing cations, carbon, boron, oxygen and fluorine. Analysis of the sublattice symmetries revealed that the carbon and oxygen sublattices in CsPbCO$_3$F, as well as the boron and oxygen/fluorine sublattices in PbB$_5$O$_7$F$_3$ and PbB$_2$O$_3$F$_2$, lack a symmetric inversion center at the unit cell, so they are in the non-centrosymmetric sublattices. In CsPbCO$_3$F, lead is constrained by the symmetry of surrounding oxygen and fluorine atoms, and their lone-pair electron show stereochemical inactivity. The F \textit{p}$_z$ orbital connected to Pb is located on the centrosymmetric sublattices (\textit{P}6/\textit{mmm}), and the interaction orbital between fluorine and cesium located in the non-centrosymmetric sublattices (\textit{P}$\bar{6}$\textit{m}2), so their possible SHG contributions may be different. The covalent interaction of the Pb 6\textit{s} and O 2\textit{p} orbitals makes lead-oxygen located in the non-centrosymmetric sublattice in PbB$_5$O$_7$F$_3$ and PbB$_2$O$_3$F$_2$, and asymmetric lobe of Pb 6\textit{s} requires O 2\textit{p} to participate in the formation. We investigated the symmetry of the electronic structure of the [PbO$_n$F$_m$] polyhedral unit of the material at the molecular orbital level (Fig. \ref{fig:homo}). The electronic wave function analysis is based on multiwfn software \cite{R46}. Both the highest occupied molecular orbital and electron local function (ELF) diagrams confirmed that the lone pair electron activity follows the order: PbO$_3$F$_6$ ${>}$ PbO$_6$F$_3$ ${>}$ PbO$_6$F$_2$, corresponding to PbB$_2$O$_3$F$_2$, PbB$_5$O$_7$F$_3$ and CsPbCO$_3$F respectively.

\begin{figure*}[!t]
	\centering
	\includegraphics[width=0.833\textwidth]{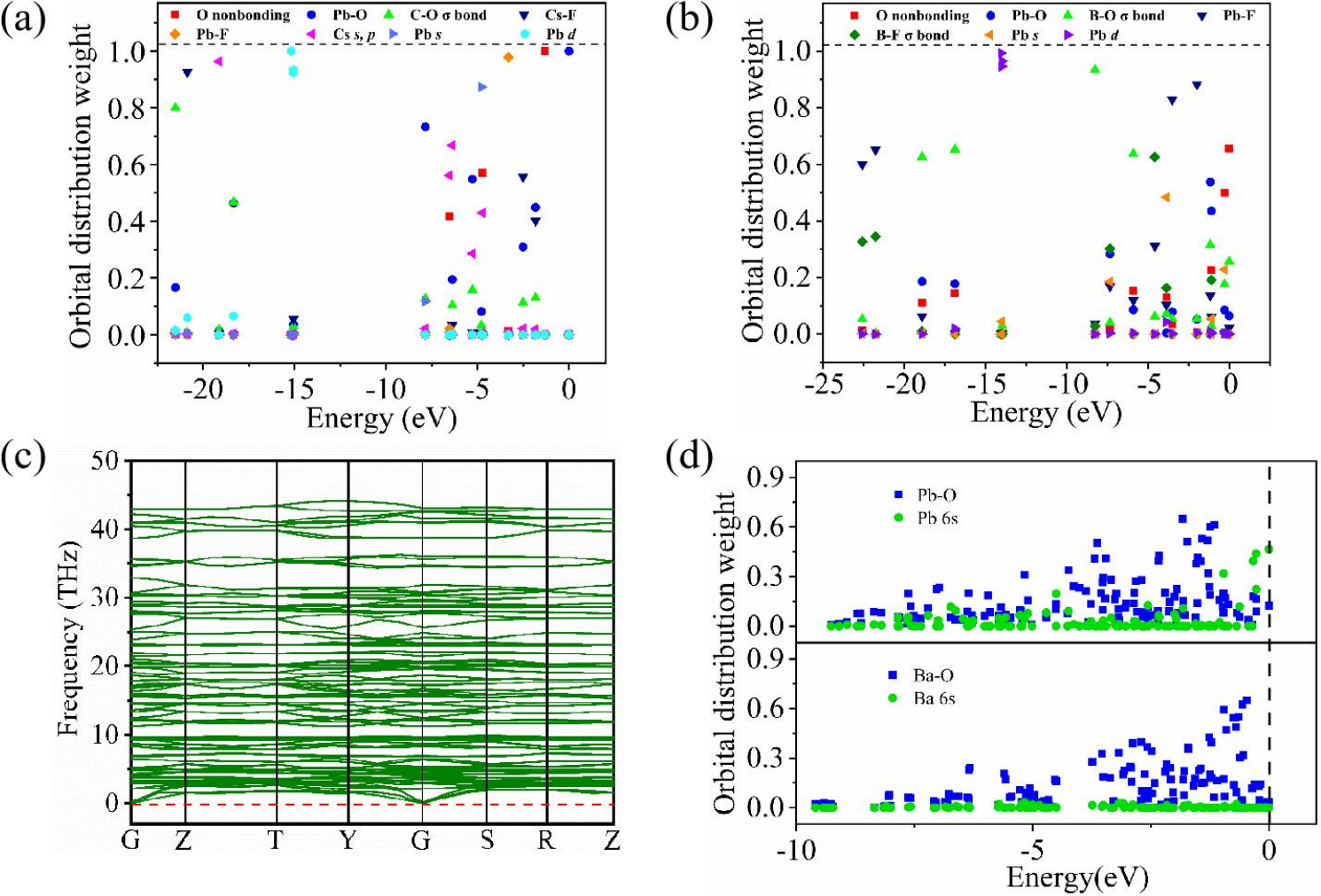} 
	\caption{(Color online) The orbital distribution weight at the energy level at \textit{$\Gamma$} point of (a) CsPbCO$_3$F and (b) PbB$_2$O$_3$F$_2$. (c) Phonon spectrum curve of BaB$_5$O$_7$F$_3$ after lead replaced barium for PbB$_5$O$_7$F$_3$. (d) The orbital distribution weight of Pb/Ba-O, Pb 6\textit{s} and Ba 6\textit{s} for PbB$_5$O$_7$F$_3$ and BaB$_5$O$_7$F$_3$.
	}
	\label{fig:orbital_weights}
\end{figure*}

To quantitatively determine the orbital contributions to the SHG coefficient from specific sublattices, we employed the symmetry-adapted WFs method. In this approach, the local orbital directions are strictly defined according to both the sublattice symmetry and coordination environment (Fig. S2). The lattice parameters of DFT and the experiment are shown in Table S1. The well consistency of the interpolation Wannier and DFT band structures are shown in Figures. \hyperref[fig:band_structure]{\ref{fig:band_structure}(a)--\ref{fig:band_structure}(c)}, indicating that the orbitals we defined are well localized.	The calculated SHG coefficients for these materials were given in Table S2. By comparing the magnitude of experimental powder SHG effect, it is demonstrated that our calculation results are reliable. For CsPbCO$_3$F [see Fig. \hyperref[fig:band_structure]{\ref{fig:band_structure}(d)}], the SHG contribution of the Pb 6\textit{s} orbitals was calculated and found to be insignificant (-2 $\%$) to the SHG. In contrast, the asymmetry of the Pb 6\textit{s} orbital in PbB$_5$O$_7$F$_3$ and PbB$_2$O$_3$F$_2$ [see Fig. \hyperref[fig:band_structure]{\ref{fig:band_structure}(e)}] indicates that lone-pair electron is stereochemical activity and contributes more to SHG (6 $\%$ and 10 $\%$). It is worth noting that the interactions orbitals between lead and oxygen contribute significantly to the SHG, with 54, 29 and 34 $\%$ for CsPbCO$_3$F, PbB$_5$O$_7$F$_3$ and PbB$_2$O$_3$F$_2$, respectively. This means that in CsPbCO$_3$F, \textit{p}-\textit{$\pi$} hybridization can induce a strong SHG effects. In PbB$_5$O$_7$F$_3$ and PbB$_2$O$_3$F$_2$, the interaction between cation and anion groups makes an important contribution to the SHG effects, while in more significant lone pair electron activity materials, the lead-oxygen orbitals may contribute even more. On the other hand, the O nonbonding orbital also contribute significantly to the SHG, with 25, 56 and 17 $\%$ for CsPbCO$_3$F, PbB$_5$O$_7$F$_3$ and PbB$_2$O$_3$F$_2$, respectively. In comparison, C/B-O \textit{$\sigma$} covalent bond contribute modestly. For fluorine-related orbitals, its contribution to the SHG is not negligible. Among them, the contributions of interactions orbitals between fluorine and lead/cesium in CsPbCO$_3$F are 3 and 29 $\%$, respectively, which are consistent with the results of symmetry analysis. Among the atoms in non-centrosymmetric sublattice, the contribution to SHG was determined by their distribution in the energy space. Therefore, we need to analyze the SHG contribution of sublattice on orbitals in energy space.

To more clearly observe the distribution of the orbital in the energy space, the SHG coefficients calculated at \textit{$\Gamma$} points is used as a reference for the clarity for CsPbCO$_3$F and PbB$_2$O$_3$F$_2$. Figures. \hyperref[fig:orbital_weights]{\ref{fig:orbital_weights}(a)--\ref{fig:orbital_weights}(b)} show the orbital distribution weight at the \textit{$\Gamma$} point of CsPbCO$_3$F. The interaction orbital between lead and oxygen, and oxygen nonbonding orbital have a substantial weight at the band edges. Moreover, the ionic bonds between fluorine and cesium for CsPbCO$_3$F, and the Pb 6\textit{s} orbital for PbB$_2$O$_3$F$_2$ also have a clear distribution at the band edges. These orbitals dominate the contribution of the SHG coefficient. The orbitals of $\sigma$ bonds between C/B-O/F are mainly distributed in the deep-level regions, and contribute weakly to SHG. Here, we constructed the isomorphic BaB$_5$O$_7$F$_3$ by using barium instead of lead to change the distribution of the band-edge orbitals of PbB$_5$O$_7$F$_3$ to illustrate the role of stereochemically active lead in the SHG effects. BaB$_5$O$_7$F$_3$ is dynamical stable [see Fig. \hyperref[fig:orbital_weights]{\ref{fig:orbital_weights}(c)}] and its shortest phase-matched wavelengths (184 nm, Fig. S4) is blue-shifted compared to PbB$_5$O$_7$F$_3$ (240 nm). We analyzed the orbital distribution weights of Pb/Ba-O, Pb 6\textit{s} and Ba 6\textit{s} in the energy space [see Fig. \hyperref[fig:orbital_weights]{\ref{fig:orbital_weights}(d)}]. The barium instead of lead causes blue-shift of band gap, which is not conducive to electron transition and thus reduces SHG. Calculation results also show that compared to PbB$_5$O$_7$F$_3$, the calculated SHG coefficient of BaB$_5$O$_7$F$_3$ are decreased sharply (decreased by 63 $\%$). The Ba 6\textit{s} orbitals are almost spherical compared to the asymmetric distribution of the Pb 6\textit{s} orbitals in PbB$_5$O$_7$F$_3$, and contribute almost zero to the SHG. Therefore, lead plays an indispensable role in the origin of the SHG, which arises from orbitals of cation and anion group interactions on the non-centrosymmetric sublattice at the band-edge.
\vspace{1\baselineskip}
{\section{CONCLUSION}}
In conclusion, we studied the origin of SHG in a series of non-centrosymmetric materials containing lone-pair electrons by analyzing the orbitals on the sublattice. A decision-tree framework was developed to systematically evaluate the influence of lone-pair electrons on SHG effect, with PbO, PbS and \textit{\text{$\beta$}}-SnF$_2$ serving as model systems to establish the hybridization interaction-SHG effect relationships. The study finds that the large SHG effects in these materials are all related to the orbitals of anionic groups. Through electron structure symmetry analysis and orbital-resolved SHG contribution calculations in CsPbCO$_3$F, PbB$_2$O$_3$F$_2$ and PbB$_5$O$_7$F$_3$, it is shown that strong stereochemically active cation-anion interaction orbitals of lone-pair electrons may contribute more to the SHG. For CsPbCO$_3$F, the SHG contribution of the Pb 6\textit{s} orbital is not significant (-2 $\%$). The asymmetry of Pb 6\textit{s} orbitals in PbB$_5$O$_7$F$_3$ and PbB$_2$O$_3$F$_2$ contributes more to SHG (6 $\%$ and 10 $\%$). To illustrate the role of stereochemically active lead in SHG effects, we substituted barium for lead, which changes the distribution of the band-edge orbitals in PbB$_5$O$_7$F$_3$. The phase matching of BaB$_5$O$_7$F$_3$ has a significantly blue-shifted (184 nm), and SHG reduced compared to PbB$_5$O$_7$F$_3$. The orbitals of the interaction between lead and oxygen are also very significant because these orbitals are located at the band edge and in non-centrosymmetric sublattices.
\vspace{1\baselineskip}
{\section*{ACKNOWLEDGMENTS}}
This work was supported by the National Key R$\&$D Program of China (2021YFB3601502), Strategic Priority Research Program of the Chinese Academy of Sciences (XDB0880000), the National Natural Science Foundation of China (22193044, 22335007, 52002397 and 22361132544), Tianshan Basic Research Talents (2022TSYCJU0001), and Chinese Academy of Sciences Project for Young Scientists in Basic Research (YSBR-024).

\end{document}